# VulCatch: Enhancing Binary Vulnerability Detection through CodeT5 Decompilation and KAN Advanced Feature Extraction

Abdulrahman Hamman Adama Chukkol[1]{ahchukkol@gmail.com}, Luo Senlin[1], Kashif Sharif[1], Haruna Yunusa[2], Muhammad Muhammad Abdullahi[3]
[1]Beijing Institute of Technology, [2]Beihang University, [3]American University of Nigeria

*Abstract*— **Binary programs vulnerability detection is a major problem in software security. Existing deep learning systems mostly rely on source code analysis for training and testing, these approaches depend on manually defined patterns, limiting their capability to detect unknown vulnerabilities. Also, the differences between models understanding of source code and compilers interpreting it into executable binaries leads to missed multiple types of vulnerabilities. Furthermore, the presence of redundant instruction set in source code complicates the training procedure and reduces the effectiveness of traditional models. To solve these problems, we proposed VulCatch, a binary-level vulnerability detection framework. VulCatch leverages a novel Synergy Decompilation Module (SDM) and Kolmogorov-Arnold Networks (KAN). By combining the power of CodeT5, it transforms raw binary code into pseudocode, thus preserving high-level semantic features. This pseudocode serves as the basis for traditional decompilers such as Ghidra and IDA to perform deep disassembly and structural analysis, while KAN enhances feature transformations and captures complex functional relationship within the code, helping in deeper analysis and recognition of underrated vulnerability patterns. This allows for effective analysis and avoids dependence on source code and addressing unknown vulnerabilities via program slicing. Code segments are transformed into vector representations by word2vec, followed by normalization. Enhancing the system more, VulCatch combines sophisticated architectural components like Inception Blocks for multi-level feature extraction from vectors, considering hierarchical and contextual code segments information, and detect multiple vulnerability types in real time using BiLSTM Attention and Residual connections on seven CVE datasets. VulCatch significantly increases detection accuracy (98.88%) and precision (97.92%) while decreasing false positives (1.56%) and false negatives (2.71%) compared to benchmarked systems.**

Abdulrahman Chukkol obtained his Masters in Information and Communication Engineering and is currently a PhD student at Beijing Institute of Technology, his research interests are Software security, Vulnerability Detection, Deep learning and neural networks.

Senlin Luo is the deputy director, laboratory director, and a professor with the Information System and Security and Countermeasures Experimental Center, Beijing Institute of Technology, Beijing 100081, China. His research interests include machine learning, medical data mining, and information security. Luo received a Ph.D. in computer science from the Beijing Institute of Technology.

Kashif Sharif received his M.S. degree in information technology and Ph.D. degree in computing and informatics from the University of North Carolina at Charlotte in 2004 and 2012, respectively. He is currently an associate professor at Beijing Institute of Technology. His current research interests include wireless and sensor networks, network simulation systems, software defined and data center networking, ICN, and the Internet of Things. He is a member of ACM.

Yunusa Haruna receives his PhD from Beihang University in China, his research interests include Pattern recognition and intelligent system, object detection, convolutional neural network, and robust computer vision algorithms.

Muhammad is a Masters student at American University of Nigeria, his research interests include, Computer security and countermeasures.

## I. INTRODUCTION

SOFTWARE vulnerability detection is crucial for enhancing cybersecurity as the frequency and severity of security incidents continue to escalate. A major challenge is the detection of clone vulnerabilities, which propagate across the software supply chain, often exacerbated by the unavailability of source code for binary programs. Traditional techniques like taint analysis, symbolic execution, fuzzing [1], and static analysis are commonly employed but face limitations in detecting unknown vulnerabilities and rely heavily on code similarity thresholds, leading to high false positive rates.

Only known vulnerabilities (cloning vulnerabilities) can be detected using code similarity-based approaches [2], it is difficult to recognize unknown vulnerabilities (recurring vulnerability), fresh vulnerabilities containing the same codes properties as the known vulnerabilities. Additionally, thresholds must be defined in order for code similarity-based detection approaches to detect whether or not the targeted functions are identical. For instance, "diff" [3], put out the top five functions that are similar to known vulnerability; "vulSeeker [2], and "SAFE" [4], also use a threshold to find similar vulnerabilities. However, setting a fixed similarity threshold can result in a high rate of false positives because functions identified as similar to known vulnerabilities may only partially match, or share superficial similarities, leading to non-vulnerable functions being mistakenly flagged, which will ultimately take a lot of time to manually examine the matching results.

Vulnerability detection techniques like, [5], based on vulnerability patterns may produce a high rate of false-positive as well as false-negative since the vulnerability features extraction is either too general or manually obtain by experts. Furthermore, extracting the features takes a lot of work. Deep learning has recently also been used to automatically extract patterns for detection of vulnerabilities. For instance, in order to extract vulnerability features more, [6] and [7], acquired the semantic information of each instruction. A hierarchical attention network is suggested by HAN-BSVD [8], to extract vulnerability features from assembly instructions. Vulnerability patterns [5], [6], [9], [10], [8], are derived from binary program assembly code or intermediate representations (IR), which are subject to the effect of compilation variability. As a result, the code representation suffers significant modifications, including reallocating registers, rearranging instructions, substituting instructions, changing the code structure, and more. These code alterations will also have an impact on how vulnerability patterns are extracted. The majority of vulnerability detection techniques [2], [3], [4], are also coarse-grained since the whole function is used to extract features. The usefulness of the extracted feature is usually impacted by other unrelated function instructions, although vulnerabilities are usually only linked to a limited number of instructions. Because compiler diversity results in considerable code changes whereas patch code only differs slightly from vulnerable code, it is particularly vulnerable to compilation diversity and patch code.

CodeT5 is an encoder-decoder model based on the T5 architecture, pre-trained on a vast code corpus to effectively interpret and generate code through tasks like summarization and translation [11]. Its ability to understand code semantics makes it valuable for code analysis and decompilation. LLM4Decompile [12] explores using large language models (LLMs) for decompiling binary code, aiming to benchmark re-compilability and re-executability. Similarly, LmPa [13] combines an LLM with program analysis to iteratively recover variable names from binary executables, enhancing readability. Challenges in decompilation, such as recovering high-level constructs from low-level assembly due to information loss during compilation, are discussed, highlighting the promise and difficulty of achieving fully recompilable decompilation with LLMs [14]. CodeT5 [15], designed for both code understanding and generation, is suitable for translating between different levels of code abstraction. It introduces identifier-aware pre-training to handle identifiers crucial for code semantics, beneficial for decompilation requiring semantic understanding. Its multi-task learning capability and broad evaluation on code-related tasks suggest its potential for integration with traditional decompilers like IDA and Ghidra. Additionally, automating the conversion of agile user stories into pseudocode is surveyed by [16], presenting a two-stage process using the CodeT5 model. This approach, fine-tuned for success measures using BLEU scores, aims to streamline software development in Agile environments by reducing the time spent converting user requirements into pseudocode, outperforming rule-based methods.

Kolmogorov-Arnold Networks (KAN) are based on the Kolmogorov-Arnold theorem, established by Vladimir Arnold and Andrey Kolmogorov in 1957. This theorem demonstrates that functions of several variables can be expressed as a superposition of functions of a single variable. KANs incorporate these univariate functions into their structure, providing an elegant method for approximating complex multivariate functions. Study [17] shows that KANs outperform traditional MLPs in capturing complexities in large datasets, offering better interpretability and accuracy. This makes KANs particularly suitable for tasks like vulnerability detection, where high-dimensional data and subtle nonlinear patterns are involved, overcoming the limitations of traditional neural networks due to the curse of dimensionality. KAN can effectively break high-dimensional linkage through the simpler one-dimensionality, linking to complex and more fine-grained data types among multiple spaces of knowledge representation at runtime point for dataset interpretability enhancement; boost detection power in a broader range with fewer perturbations of discoverable target functions. $f$ is the multi-variate function to be signified. $\Phi_q$ and $\emptyset_{q,p}$ are uni-variate functions. $x_1, x_2, \ldots, x_n$ are the variables of function $f$.

$$f(x_1, x_2, \ldots, x_n) = \sum_{q=1}^{2n+1} \Phi_q \left( \sum_{p=1}^{n} \emptyset_{q,p}(x_p) \right) \quad (1)$$

This formula shows that function $f$ of $n$ variables can be decomposed into a superposition of $2n + 1$ functions $\Phi_q$, each of them depends on a particular sum of functions $\emptyset_{q,p}$, letting each $\emptyset_{q,p}$ are functions of a distinct variable $x_p$. Here, $\Phi_q$ represents the outer functions, and $\emptyset_{q,p}$ represents the inner functions, each applied to one of the variables. The inner functions transform the individual variables, and their results are then summed up as inputs to the outer functions. In (section 3), VulCatch's combined dataset $C$ analogous to $x_1, x_2, \ldots, x_n$ is processed by the KAN function $F_{KAN}$. This function embodies the KAN formula where each component of the dataset is treated as an input variable $x_p$ to the functions $\emptyset_{q,p}$. These are then aggregated (summed) and passed through the functions $\Phi_q$ to generate enhanced features $K_{features}$.

Decompilation technology has advanced rapidly [18], with tools like IDA Pro [19] and Ghidra [20]. Many binary code analysis fields require decompilation techniques [21], often using intermediate representations (IR) as in Inceptions and FirmUp [21]. Decompiled pseudo-code is used in cybersecurity; for example, UPPC (Unleashing the Power of Pseudo-code) [22] uses pseudo-code for vulnerability detection due to its higher abstraction and platform independence. Other studies use pseudo-code to counter compilation diversity [23] and detect malware via decompiled Android API calls. SCRUTINIZER [24] constructs pseudo-code from memory dumps to detect malicious code repetition. Decompiled pseudo-code offers high-level semantic information and control structure recreation, making it more efficient for binary security analysis compared to analyzing binary code or IR.

Our primary research goal is to develop a system capable of detecting multiple types of both known and unknown vulnerabilities in binary code more accurately and efficiently than current methods. The specific research questions we address are:

VulCatch begins by decompiling binary code into pseudocode using the CodeT5 model, making it more understandable for further analysis. Disassembly analysis with IDA and Custom Analysis, along with structural analysis using Ghidra and Control Path, provides detailed insights into the code's structure and behavior. The data from these analyses, combined with metadata, is enhanced using Kolmogorov-Arnold Networks, improving the feature set. Abstract Syntax Trees (ASTs) and Program Dependence Graphs (PDGs) are generated from these features, forming structured representations of the code. These are prepared as inputs for a BiLSTM-Residual-Attention Neural Network, with residual connections aiding in training deeper networks. Bidirectional passes in the BiLSTM capture context from both past and future states, while inception block processing and an attention mechanism focus on important parts of the input sequence. Residual connections enhance gradient flow, and the deep learning model classifies vulnerabilities from the processed input data. The final output translates these classifications into actionable information, identifying vulnerabilities with their categories and severity levels, providing valuable insights for remediation.

Experimental results demonstrate VulCatch's superior performance across various datasets. On the SARD-AdvOpt-SDM dataset, it achieves an accuracy of 98.43%, precision of 97.92%, and an F1-score of 98.88%. On the SARD-TestVul-SDM dataset, VulCatch attains an accuracy of 99.29%, precision of 99.50%, and an F1-score of 99.05%. Testing on the SARD-CWE-HighOpt-SDM dataset shows an accuracy of 99.19%, precision of 99.45%, and an F1-score of 98.90%. On the SARD-CWE-LowOpt-SDM dataset, it achieves an accuracy of 98.79%, precision of 99.00%, and an F1-score of 98.88%. This consistent performance across multiple datasets highlights VulCatch's effectiveness in detecting a wide range of vulnerabilities.

The key contributions of this research are:

- We present Synergy Decompilation Module (SDM), a comprehensive analysis on advanced decompilation techniques, leveraging CodeT5 integrated with traditional decompilers (Ghidra and IDA), enhancing the accuracy and completeness of binary pseudocode representation for vulnerability detection.
- We integrated Kolmogorov-Arnold Networks (KANs) to enhance feature transformation capabilities, which results in outperforming traditional neural network models in terms of detecting unknown vulnerabilities, the integration of KAN reduces false positive rates in binary vulnerability detection.
- We implemented program slicing techniques to refine vulnerability detection and developed VulCatch, a novel system consisting of BiLSTM, attention mechanism, and residual connections for real-time vulnerability detection.
- We curated a diverse set of datasets used to evaluate VulCatch's performance across different optimization levels and scenarios, demonstrating its robustness and adaptability to varying compilation strategies.

This paper starts by introducing the research background, significance, and problems with current methods, followed by the research goals and proposed solutions in *section 1*. It then reviews related work on vulnerability patterns and code similarities in *section 2*. The paper describes the system components in *section 3*. It details the experimental procedure, neural network design, dataset construction, evaluation measures and analysis of results in *section 4*. Finally, the paper concludes with a discussion, followed by the limitations and future work in *section 5*.

## II. RELATED WORKS

In the evolving field of binary vulnerability detection, two main methodologies are often applied: code similarity-based approach and vulnerability pattern extraction approach. Code Similarity-Based Techniques: These methods compare target codes to databases of known vulnerabilities to measure their similarities. The target code is considered vulnerable when a predetermined threshold is exceeds. Nevertheless, the threshold can cause high false-positive rates due to the superficial similarities. Vulnerability Pattern Extraction: This method identifies property signatures or patterns to detect known vulnerabilities in target code. By comparison against predefined patterns derived from previous vulnerabilities, it then flags codes with identical features. However, it is limited to known patterns and might not understand new vulnerabilities. Limitations and Challenges: Both methods struggle with false positives and variability, mainly due to compiler differences and irrelevant code throughout the functions. Current research aims to solve these problems through advanced decompilation and deep learning for accurate and efficient vulnerability detection.

### A. Approaches based on Vulnerability patterns

Similarity based techniques are used to find recognized vulnerability (replica vulnerability). Using same approach, they can barely discover unknown vulnerability. The approaches derived from vulnerability patterns obtain the appropriate features of the current vulnerable program, using such patterns to locate the target code. IntScope, decompiles the binaries into an intermediary representation and then obtains important properties via symbol and taint approaches for vulnerability detection. With the use of software slicing and symbol execution approaches, Firmalice [31], extracts properties to suggests unique models to find concealed verification avoidance backdoors for focused embedded systems. Value set analysis is used by GUEB, to track the usage and releases of memories throughout programs in order to find binary vulnerability in UA. Pattern matching is used by cwe checker [25], to find unknown vulnerabilities using CWE vulnerability patterns provided by human analysts. These efforts, nevertheless, have large false-positive rates since the patterns and instructions are very broad and some properties are physically obtained by humans. The results must also be carefully verified. Deep learning techniques are used by VulCatch to automatically extract vulnerability models.

Several researches then apply deep learning to automatically detect vulnerabilities patterns to address the existing challenges. Instruction2Vec [6], creates a vector representation of the instruction through dividing instruction as opcode and operands (having two features), each property representing a vector. Trained Text-CNN applies feature extraction of weak code using the learnt vector representation. According to [7], proposal, every instruction should be presented in a meaningful combination given by its vector element, with a weighted summation instruction applied according to the element's hierarchy level. The BiLSTM system is then trained to recognize patterns of vulnerabilities. According to the binary program machine instruction structure, DCKM [26], combines Cost Sensitive Kernel Mechanism and BRNN in finding vulnerabilities in binaries of various feature spaces. HAN-BSVD [8], builds a network for extracting features to catch spatial information that indicate critical locations for collecting important vulnerability features by first training their instruction embedding models consisted of BiGRU and word-attention modules. Zheng, suggested converting binary codes into LLVR IR and collecting instructions relating to library functions using data flow analysis. Eventually, those instructions are used to train various RNN networks for vulnerability findings. VulDeePecker [27], a model for detecting vulnerabilities in source code, is adopted by BVDetector [9], but it collects code slices associated with API and library function at assembly level. Such approaches were centered on low-level code properties, making granular analysis become finer. VulCatch extracts vulnerability patterns using a finer-grained method. Patch-code is used in other techniques for feature extraction. For instance, VIVA [10], efficiently detects known and unknown vulnerabilities by using binary program slicing methods along with decompilation to provide vulnerability and patch patterns. Such approach, meanwhile, is costly and subject to compilation complications. VulCatch simply extract features relating to the detection of vulnerabilities.

*B. Approaches based on Code similarity*

Code similarities have been extensively used in malware classification [28], plagiarism detection [29], and security patch assessment [30], as well as vulnerability findings [34]. In order to identify binary vulnerability that are same as existing vulnerabile code, TEDEM, introduced tree distance control quantifying similarities in codes at the level of basic blocks. In order to pre-filter the target function, DiscovRE, collects numerical information from the function (such as the quantity of mathematical and analytical process). Similarities between the structure's features are then analyzed in order to identify vulnerabilities. Dependent formulae are semantic properties extracted at high-level to detect vulnerable codes XMATCH [32], from the original binary code. Such vulnerability detection techniques, though, frequently generate expensive similarity calculations and have high false positive and false negative rates. Machine learning methods have been applied to these issues to compare code similarity and find vulnerabilities. Genius [33] creates attributable CFGs and determines similarities using graph embeddings produced by comparison with a collection of sample graphs known as codebook. Gemini is a GNN-based model that was proposed by [34], that uses structure2vec, in combination with Siamese model, to produce the function graphs embedding and determine match. In order to enhance the detection accuracy, VulSeeker [35], builds the labelled semantics flow graphs (LSFG) using data control in functions with embedding LSFG in the space vectors via DNN semantical awareness framework.

However, such techniques depict fundamental building blocks or functions using certain manually chosen statistical features that might not have adequate semantic information. adiff [36], extracts three categories of semantic information for similarity detection from introduced functions, raw bits functions, and function calls. In order to identify potential vulnerability functions, FIT [37], first extracts semantic information via neural network framework for binary programs. Moreover, it further analyzes the similarity of two features based on third-level features. With the help of assembler instructions, SAFE [4], collects semantic information. Next, using a transformer, it creates embeddings of functions. In order to identify vulnerabilities, it performs similarity calculations on the embedded representations that were developed.

Several studies propose the introduction of patch codes to find vulnerabilities in order to lower the rates of false positive more. In order to create a trace set, BINXRAY [38], first identifies the fundamental building elements that are different between the patch and vulnerable functions. In order to determine patch in the functions, such checks help in computing similarities among patch, target and vulnerable functions accordingly. PDiff [39] compared the kernel's target before the adoption and after the adoption of patch with its normal versions and produces a description with semantics linked to the targeted patch. Lastly, every target program's patch status is determined using the similarity.

Table 1 Comparison of the Related Works

| Reference | Technique | Approach | Description | Strengths | Limitations |
|---|---|---|---|---|---|
| [25] | CWE Checker | Vulnerability Pattern | Uses pattern matching for CWE patterns. | Detects unknown vulnerabilities. | High false-positive rates. |
| [6] | Instruction2Vec | Deep Learning for Vulnerability Model | Creates vector representations, uses Text-CNN. | Detailed instruction representation. | Computationally intensive. |
| [26] | DCKM | Deep Learning for Vulnerability Model | Combines Cost-Sensitive Kernel Mechanism and BRNN. | Handles various feature spaces. | Complexity in model training. |
| [8] | HAN-BSVD | Deep Learning for Vulnerability Model | Uses BiGRU and word-attention modules. | Effective feature extraction. | High computational cost. |
| [9], [27] | VulDeePecker | Deep Learning for Vulnerability Model | Adapts for binary code, focuses on API/library functions. | Finer-grained analysis. | Limited to specific vulnerabilities. |
| [10] | VIVA | Patch-Code for Feature Extraction | Uses binary program slicing and decompilation. | Effective for known/unknown vulnerabilities. | Costly, compilation issues. |
| [31] | Firmalice | Vulnerability Pattern | Uses software slicing and symbolic execution for backdoors. | Detects backdoors effectively. | High false-positive rates. |
| [32] | XMATCH | Code Similarity | Extracts formulas and semantic properties. | High-level semantic extraction. | Expensive calculations. |
| [33] | Genius | Code Similarity | Creates attributable CFGs, uses graph embeddings. | Effective graph-based comparison. | High false-positive/negative rates. |
| [34] | Gemini | Code Similarity | Uses structure2vec and Siamese model for function graph embeddings. | Effective function matching. | High computational cost. |
| [35] | VulSeeker | Code Similarity | Builds labeled semantic flow graphs, embedding via DNN. | Improved detection accuracy. | High computational cost. |
| [36] | adiff | Code Similarity | Extracts semantic information from functions and calls. | Detailed multi-level semantic analysis. | High computational cost. |
| [37] | FIT | Code Similarity | Analyzes feature similarity at three levels. | Multi-level feature analysis. | High computational cost. |
| [4] | SAFE | Code Similarity | Collects semantic info using assembler instructions, creates embeddings. | Effective semantic information collection. | High computational cost. |
| [38] | BINXRAY | Patch-Code for Similarity | Identifies differences between patched/vulnerable functions. | Effective patch detection. | Requires pre-existing patch info. |
| [39] | PDiff | Patch-Code for Similarity | Compares kernel before/after patch adoption. | Detailed semantic patch analysis. | High computational cost. |

Code similarity-based techniques, though, can only discover clone vulnerabilities; they cannot find recurring vulnerabilities. Furthermore, in order to compare the outcomes of this approach, numerous known vulnerability databases must be built beforehand, the database's completeness will also have an impact on the detection outcomes. Without the requirement to create a database of known vulnerabilities, VulCatch may concurrently detect clone and recurrence vulnerability. The comparison of the related works is shown in *Table 1*.

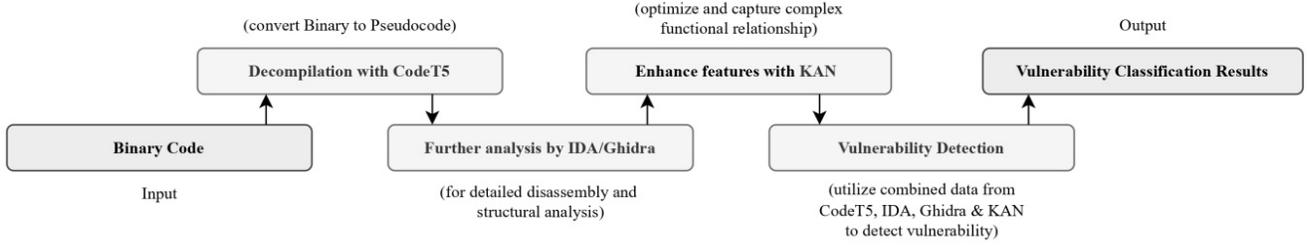

Fig. 1. Synergy Decompilation Module with KAN

### III. DESIGN OF VULCATCH

*A. System overview*

Binary code vulnerability detection is a complicated process that require complex mechanisms and methods. Traditional models mainly have problems because they are unable to understand the complex structures and semantics information embedded in binary codes. The VulCatch network solves this problem by innovating a combination of advanced decompilation procedures and deep learning framework to improve the detection of both known and unknown vulnerabilities.

The different components are explained below:

Decompilation with CodeT5: The VulCatch system starts the vulnerability detection process with CodeT5, a model trained to decompile binary code into pseudocode. This step is needed because it transforms raw binary into a more interpretable format and preserves high-level semantics properties. The pseudocode generated is highly rich in semantic information, critical for the understandings of complex software behaviors that might hide vulnerabilities *Fig. 1* Novel SDM framework.

Traditional Decompilers Integration: Following the decompilation stage, traditional tools Ghidra and IDA are applied for detailed disassembly and structural analysis. The tools complement CodeT5 by maintaining a granular breakdown of the binary structures, which is essential for the comprehensive examination of vulnerabilities *Fig. 1*.

KAN is placed before the feature extraction phase and after the initial data aggregation phase. This position allows KAN to process the raw combined data from disassembly, improving representation of the data before it is used to generate more structural Abstract Syntax Trees (ASTs) and Program Dependency Graphs (PDGs).

Abstract Syntax Trees (ASTs) and Program Dependency Graphs (PDGs): The model then generates ASTs and PDGs from the combined data of CodeT5's pseudocode and traditional decompiler output. ASTs benefit in understanding the hierarchical structure of the codes, while PDGs concentrate on the interactions and dependencies among various elements. These structures are helpful in pinpointing critical parts of the code that are more likely to have vulnerabilities.

Deep Learning Techniques for Enhanced Detection: To analyze and process data from ASTs and PDGs, VulCatch uses a BiLSTM network integrating attention network with residual connectivity in *Fig. 2* VulCatch system design. This system is adept at detecting complex dependencies and patterns that are indicative of vulnerabilities. The attention mechanism precisely allows the model to focus on portions of the code that have greater probabilities of having vulnerabilities, while residual connection enhances flow of gradient, helping effective learning and improved generalization, thus enhancing the efficiency and accuracy of the detection process.

Inception blocks are integrated into the deep learning structure to improve feature extraction by letting the network capture multi-scale patterns in the data. These blocks contain multiple convolutional layers with different filter sizes that operate in parallel. The inception blocks assist in detecting vulnerabilities at different levels of granularity, thus increasing the robustness of the detection system.

The VulCatch framework is built to improve binary vulnerability detection by integrating advanced tools and methods. The method is organized around a sequence of procedures that leverages the exclusive strengths of CodeT5, traditional decompilers IDA and Ghidra, KAN, and current deep learning methods.

**Algorithm 1:** The Synergy Decompilation Module, starts by transformation of BinaryCode into pseudocode via DECOMPILE_WITH_CODET5, to make sure the semantic integrity of the original code is well-kept. The pseudocode is disassembled using IDA to enhance its original structure, and any structural or logical issues are found by additional analysis using Ghidra. Ghidra and IDA's outputs are combined to construct a single, comprehensive dataset. Kolmogorov-Arnold Networks (KAN) now processes this dataset further, by the use of a nonlinear mapping to transform features and refine them, which is important for possible vulnerability detection. After this process, the enhanced datasets are then used to generate ASTs and PDGs. These structures aids to visualize the program's hierarchical and dependent aspects essential for detection and categorization of vulnerabilities according to their nature and severity. A list of categorized vulnerabilities is produced by this process, together with

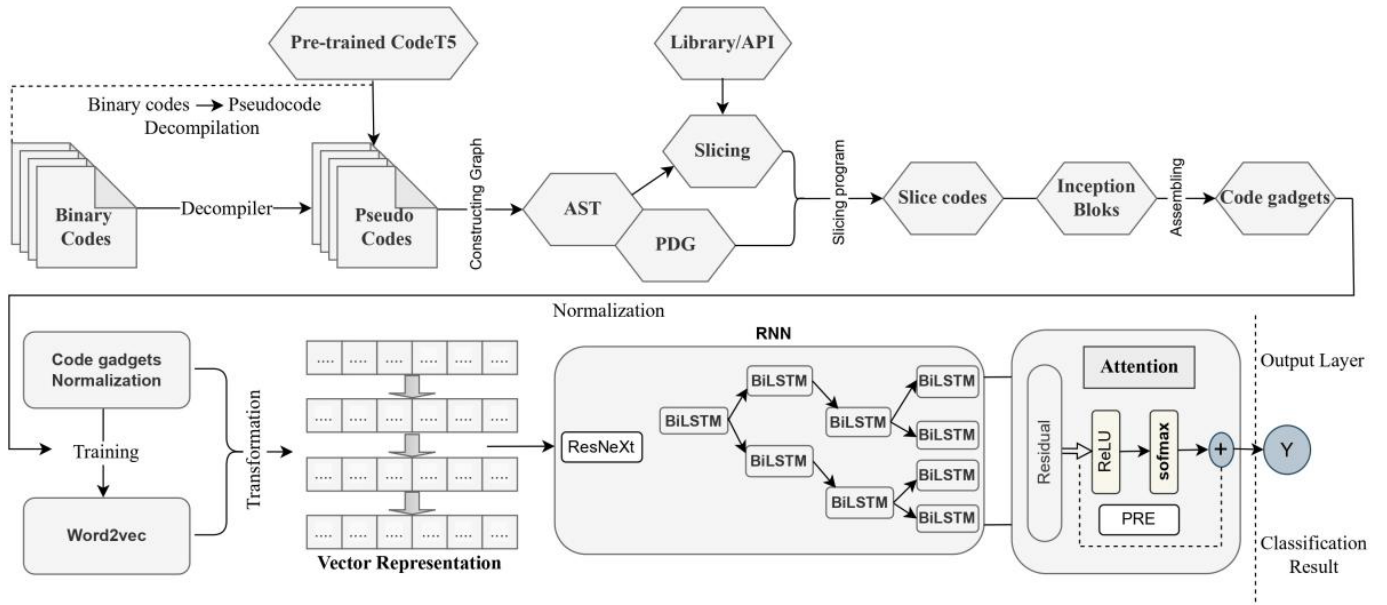

Fig. 2. Overall VulCatch system design

```
Algorithm 1 Synergy Decompilation Module & KAN

Input: BinaryCode
Output: ClassifiedVulnerabilities

1:  procedure VULCATCH(BinaryCode)
2:      PseudoCode ← DECOMPILE_WITH_CODET5(BinaryCode)
3:      DisassemblyReport ← IDA_DISASSEMBLE(PseudoCode)
4:      StructuralAnalysis ← GHIDRA_ANALYZE(PseudoCode)
5:      CombinedData ← COMBINE (DisassemblyReport StructuralAnalysis)
6:      EnhancedFeatures ← APPLY_KAN(CombinedData) // Data features enhance with KAN
7:      ASTs ← GENERATE_ASTS(EnhancedFeatures) // ASTs generation from enhanced features
8:      PDGs ← GENERATE_PDGS(EnhancedFeatures) // PDGs similarly generate from enhanced features
9:      Vulnerabilities ← DETECT_VULNERABILITIES (ASTs, PDGs)
10:     ClassifiedVulnerabilities ← VULNERABILITIES_CLASSIFICATION (Vulnerabilities)
11:     return ClassifiedVulnerabilities
12: end procedure
```

information about vulnerable binary codes and a systematic assessment.

$$PseudoCode = F_{c5}(BinaryCode) \quad (2)$$

The purposes of CodeT5 model used to decompile binary code into pseudocode, are represented by $F_{c5}$. $BinaryCode$ is the input decompiled into $PseudoCode$, which becomes the pseudocode output of the CodeT5 model.

$$D_{IDA} = F_{IDA}(PseudoCode) \quad (3)$$
$$D_{CA} = F_{CA}(PseudoCode) \quad (4)$$

(IDA & CA) Disassembly Analysis $F_{IDA}$ and $F_{CA}$ functions correspondingly signify the processes of Custom Analysis (CA) and the Interactive Disassembler (IDA) to examine the pseudocode. $PseudoCode$ serves as the input to these functions, which was obtained from the decompilation step. $D_{IDA}$ and $D_{CA}$ are the outputs of the IDA and CA analysis.

$$S_G = F_G(PseudoCode) \quad (5)$$
$$S_{CP} = F_{CP}(PseudoCode) \quad (6)$$

Structural Analysis (Ghidra & CP) takes $F_G$ and $F_{CP}$ as the functions representing the Ghidra analysis and Control Path analysis (CP), respectively. $S_G$ and $S_{CP}$ are the structural analysis outputs from Ghidra and CP.

$$C = F_{Combine}(D_{IDA}, D_{CA}, S_G, S_{CP}, M) \quad (7)$$

Data Combination $F_{Combine}$ is the function that integrates data from various analysis tools along with metadata. $M$ is the metadata that provides additional context or information for the combination process. $C$ represents the combined dataset used for further analysis.

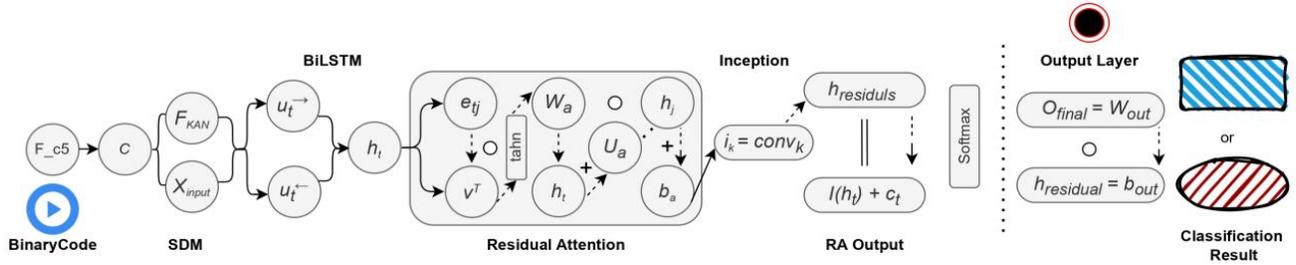

Fig. 3. VulCatch's Network

$$K_{features} = F_{KAN}(C) \quad (8)$$

$K_{features}$ is the output after applying the Kolmogorov-Arnold Networks to the combined data. $F_{KAN}$ symbolizes the application function of KAN, which is used here to enhance the feature set derived from the combined data $C$. The combined data $C$ includes outputs from various analysis tools like IDA, CA, Ghidra, and others, merged with additional metadata.

$$ASTs = F_{AST}(K_{features}) \quad (9)$$
$$PDGs = F_{PDG}(K_{features}) \quad (10)$$

AST and PDG generation uses $F_{AST}$ and $F_{PDG}$ as the functions that generate ASTs and PDGs from the combined data. $ASTs$ and $PDGs$ outputs represents structured representations of the code for vulnerability analysis.

B. BiLSTM-Residual-Attention Neural Network

This work deploys Recursive Neural Networks (RNNs), which have demonstrated superior performance in natural language processing [40], in conjunction with the sophisticated bidirectional gating mechanisms of BiLSTM [41]. This integration is tailored to capture the hierarchical relationships in data, essential for producing accurate vulnerability patterns. Our novel framework, VulCatch, incorporates a Bidirectional Hierarchical Network. This development is aimed at extracting the representations from the hierarchical nature of data. The representations produced are then used as inputs to the BiLSTM layer. Here, the BiLSTM model constructs a representation vector from the AST, effectively compressing its semantic information for vulnerability detection in *Fig. 3* VulCatch Network.

$$X_{input} = (AST_s, PDG_s) + Residual\ Connections \quad (11)$$

$AST_s, PDG_s$ are prepared for input into the deep learning model, specifically the ASTs and PDGs generated from the enhanced features $K_{features}$. $Residual\ Connections$ allows for direct connections between earlier layers and later layers, adding the residual connections to the ASTs and PDGs better propagates gradients through the deep network during training, thus alleviating the vanishing gradient problem and improving the learning of deeper layers.

$$x_t = Extracted\ feature\ at\ time\ t\ from\ X_{input} \quad (12)$$

$x_t$ is the specific feature vector extracted at time $t$ from $X_{input}$ ensuring that the BiLSTM network processes the correct sequence of enhanced data.

$$\overrightarrow{u_t} = \sigma(W_{u,f} \cdot [x_t, \overrightarrow{h_{t-1}}, \overrightarrow{c_{t-1}}] + U_{u,f} \cdot \overrightarrow{h_{t-1}} + V_{u,f} \cdot \overrightarrow{c_{t-1}} + b_{u,f}) \quad (13)$$

Forward pass is set as $\sigma$ being the sigmoid activation function that maps the calculated values to a range between 0 and 1. $W_{u,f}$, $U_{u,f}$, $V_{u,f}$ are weight matrices specific to the update gate for the forward pass of the LSTM. $b_{u,f}$ is the bias term for the update gate in the forward pass. $x_t$ is the input at time step $t$. $\overrightarrow{h_{t-1}}$ is the previous hidden state from the forward direction. $\overrightarrow{c_{t-1}}$ is the previous cell state from the forward direction.

$$\overleftarrow{u_t} = \sigma(W_{u,b} \cdot [x_t, \overleftarrow{h_{t+1}}, \overleftarrow{c_{t+1}}] + U_{u,b} \cdot \overleftarrow{h_{t+1}} + V_{u,b} \cdot \overleftarrow{c_{t+1}} + b_{u,b}) \quad (14)$$

Backward pass is set as $\overleftarrow{h_{t+1}}$ and $\overleftarrow{c_{t+1}}$ being the subsequent hidden and cell states from the backward direction. $W_{u,b}$, $U_{u,b}$, $V_{u,b}$ And $b_{u,b}$ are the weight matrices and bias for the update gate in the backward direction. This pass processes the sequence from end to start, providing a perspective on the data that complements the forward pass.

$$h_{inception} = I_k(h_t) = \sum_f Conv_k(h_t, f) \quad (15)$$

Inception Blocks takes $I_k$ as the inception block processing hidden states $h_t$ using convolutional filters of size $k$. $Conv_k$ is the convolution operation with kernel size $k$. $f$ serves as the index of the convolution filter.

Furthermore, VulCatch is enhanced with Residual connections and Attention Mechanism, which includes Attention Units that are divided into a mask and a trunk division. The trunk division is tasked with feature processing and is designed to be compatible with current network structures, such as pre-activation Residual Units [42], ResNeXt [43], and Inception models [44]. The mask division learns a same-size mask $M(x)$ that softly weights the output features $T(x)$ provided by the trunk division. This design is inspired by the fast feed forward and feedback attention processes, structuring attention in a top-down, bottom-up fashion [45], which in turn heads the trunk division's neurons

similarly to a Highway Network with the output mask [46]. The Attention weights looks like this:

$$e_{tj} = v^T \cdot \tanh(W_a \cdot h_t + U_a \cdot h_j + b_a) \quad (16)$$

Attention Mechanism's $v, W_a, U_a$ are the parameters of the attention mechanism that weight the importance of different parts of the input sequence. $h_t, h_j$ are the current and other hidden states on which attention is being computed. $b_a$ is the bias term for the attention mechanism. Tanh is the hyperbolic tangent activation function providing non-linearity.

$$h_{residuals} = I(h_t) + c_t \quad (17)$$

Residual Connection's $I(h_t)$ is the output from the inception block. $c_t$ is the context vector from the attention mechanism. $h_{residuals}$ is the combined output that adds the input $h_t$ to its transformation, promoting gradient flow and reducing the risk of vanishing gradients.

$$V_{output} = F_{DL}(X_{input}) \; Classified \; Vulnerabilities \quad (18)$$
$$F_{classify}(V_{output})$$

$V_{output} = F_{DL}(X_{input})$ outputs the deep learning model $F_{DL}$, which processes the input data $X_{input}$. $Classified \; Vulnerabilities \; F_{classify}(V_{output})$ categorizes the vulnerabilities detected by the deep learning model into specific classes based on severity, type, or other relevant criteria. It essentially translates the raw model outputs into actionable information, facilitating prioritization and remediation efforts.

The final output of the model is formulated as:

$$O\;final = W_{out} \cdot h_{residuals} + b_{out} \quad (19)$$

Final Output $W_{out}$ is the weight matrix for the final layer, translating the high-level features captured by the network into meaningful classifications. $h_{residuals}$ is the final feature representation derived from the network's residual and attention mechanisms. $b_{out}$ is the bias term to adjust the classification layer's final output. $O\;final$ is the final prediction or classification output of the model, representing the detected vulnerabilities with their assigned categories or severity levels.

VulCatch represents an incremental architecture that captures intricate details within the code's structure, emphasizing the critical segments for vulnerability detection through an advanced attention mechanism.

## IV. IMPLEMENTATION AND EXPERIMENTS

### A. Experimental procedure

The logic is to assemble library/API function of C/C++ relevant to vulnerabilities for obtaining library/API function calls program-slices. We gathered the Checkmarx 2021 [50], C/C++ vulnerability criteria. Checkmarx is a cutting-edge static source code analysis tool that can technically and logically locate, recognize, and monitor vulnerabilities in the majority of mainstream code. Checkmarx would still skip certain targets despite having strong coverage for SARD's vulnerability programs [51]. To create PDGs and AST of functions in all programs for program-slice, we study the C/C++ open-source code analysis algorithm Joern.

Decompilation software's are widely accessible nowadays, like Ghidra [20] and IDA Pro [19]. Because of the difference in quality of the pseudo codes generated by the decompilers, not all of the decompiled pseudo codes are usable. That is why we came up with the Synergy Decompilation Module where the CodeT5 model translates binary code into a high-level pseudocode, maintaining the abstract semantics. This pseudo code then acts as a foundation for traditional decompilers IDA and Ghidra, which perform in-depth disassembly and structural analysis. By providing these decompilers with semantically rich pseudo code, CodeT5 enhances their capabilities for more accurate disassembly and structural analysis. We compare the decompilers Ghidra [20], IDA Pro [19], and our SDM with the same binary programs. The decompiled pseudo codes produced by the SDM provide certain benefits for retrieving high-level semantics and are also easily readable. According to our analysis, SDM performs roughly 2% better than IDA Pro and 4% better than Ghidra when it comes to retrieving function boundaries and instructions recovery [52]. Furthermore, Ghidra has 5.10 times more decompilation problems than IDA Pro, and IDA Pro has 2.55 more decompilation problems than the SDM according to [53], who reprocessed the well-established compilers test methods known as EMI testing [54]. We used CodeT5, IDA Pro and Ghidra as the decompilation synergy to decompile the codes with KAN enhancements, the data values are shown in

*Table* 2 datasets properties.

### B. Construction of dataset

**Enhanced CVE-Based Binary Collection:** Our dataset has been meticulously curated to include binaries associated with a selected array of known CVE vulnerabilities [47], presenting a diverse spectrum of security challenges, ranging from server-side application vulnerabilities to complex microprocessor architectural weaknesses, ensuring a comprehensive analysis. This collection now exclusively comprises:

- CVE-2018-1002105, detailing privilege escalation vulnerability within Kubernetes API servers.
- CVE-2019-0708, also known as BlueKeep, which is a remote code execution vulnerability found in Windows Remote Desktop Services.
- CVE-2017-5715 and CVE-2017-5754, recognized respectively as Spectre and Meltdown, both of which expose speculative execution vulnerabilities that potentially allow unauthorized information disclosure.
- CVE-2018-8174, a vulnerability that enables remote code execution through the Windows VBScript engine.

Table 2 Properties of Selected Datasets

| Datasets | Training | Testing | Vulnerable | Non-vulnerable | Code Gadgets |
|---|---|---|---|---|---|
| CVE-Binaries-SDM | 65000 | 95000 | 70000 | 90000 | 160000 |
| - CVE-2018-1002105 | 18750 | 1250 | 15000 | 5000 | 23881 |
| - CVE-2019-0708 | 25000 | 2500 | 20000 | 7500 | 28656 |
| - CVE-2017-5715 | 15000 | 2500 | 12000 | 5500 | 19104 |
| - CVE-2017-5754 | 22500 | 2000 | 18000 | 6500 | 21492 |
| - CVE-2018-8174 | 21250 | 3750 | 17000 | 8000 | 26269 |
| - CVE-2020-0601 | 23750 | 1250 | 19000 | 6000 | 23881 |
| - CVE-2020-1472 | 20000 | 2500 | 16000 | 6500 | 16716 |
| SARD-CWE-LowOpt-SDM | 162382 | 277267 | 103654 | 162382 | 266036 |
| SARD-CWE-HighOpt-SDM | 150838 | 284163 | 114023 | 150838 | 264861 |
| SARD-AdvOpt-SDM | 83368 | 94624 | 42798 | 83368 | 126166 |
| SARD-TestVul-SDM | 6456 | 9928 | 3391 | 6456 | 9847 |

- CVE-2020-0601, identified as CurveBall or ChainOfFools, a spoofing vulnerability within the Windows CryptoAPI.
- CVE-2020-1472, dubbed Zerologon, an elevation of privilege vulnerability in the Microsoft Netlogon service.

**Source Code Acquisition and Enhanced Binary Compilation with Clang:** We procure source code with established vulnerabilities from the NIST SARD, utilizing the Juliet Test Suite [48] version 1.3 for C and C++. To prepare our binary analysis, we leverage Clang; part of the LLVM compiler suite, renowned for its potent optimization and comprehensive support for modern C and C++ standards. The programs are compiled into ELF 64-bit binaries, initially with Clang's **-O0** optimization level to maintain straightforward correspondence with the source code. For a thorough exploration of how optimization may affect vulnerability detection, we proceed to compile the source code using Clang's **-O1**, **-Os**, and **-Ofast** optimization levels. These distinct optimization tiers are judiciously chosen to challenge the robustness of our vulnerability detection model. The aim is to scrutinize its performance across a spectrum of compiled binaries, from the least to the most optimized, to validate and enhance its effectiveness in real-world scenarios.

**Enhanced Vulnerability Labeling:** Leveraging the Juliet Test Suite, we guarantee that each test case is annotated with a CWE identifier [49], which precisely indicates the type of vulnerability present. For the purposes of our study, we select a comprehensive set of 101 distinct CWE classes, reflecting a diverse and extensive range of security vulnerabilities. This selection allows us to compile binaries that embody a wide spectrum of real-world vulnerabilities, facilitated through both library and API calls. The CWE classes ensure that our dataset not only covers a broad cross-section of common vulnerabilities but also provides the granularity needed for the SDM to learn and detect nuanced patterns within the vulnerability landscape effectively. Choosing this much CWE classes offers a substantial variety of vulnerability types for our model to handle, which is likely to strengthen its detection capabilities across a range of scenarios.

**Integrative Analysis via SDM:** The SDM seamlessly bridges the gap between binary-level vulnerabilities and source-level semantics. By employing CodeT5, the SDM decompiles binaries—encompassing a diverse array of CVE vulnerabilities, compiles at varying optimization levels from the SARD dataset—to accurate pseudocode. This intermediate representation preserves the critical semantic details altered by optimizations ranging from the CVE identifiers and CWE optimization levels, a necessary step for robust vulnerability detection. The pseudocode undergoes further scrutiny with IDA and Ghidra, which elaborate the disassembly into comprehensive ASTs and PDGs.

**Training and Testing Set Formation:** From the compiled binaries, we allocate 80% for training purposes and reserve 20% for testing. This selection is randomized to ensure representative distribution across the dataset.

In our analysis, code gadgets are meticulously labeled for vulnerability detection: "0" for non-vulnerable functions and "1" for vulnerable functions, identified by harmful library or API calls. We have curated distinct datasets to precisely evaluate the performance of the SDM across various conditions:

Table 3 Selected datasets information

| Dataset | Optimization Levels | Compilation Perspective | Importance |
| --- | --- | --- | --- |
| SARD-AdvOpt-SDM | O2, O3 | Tests SDM with advanced optimizations. | Demonstrates robustness in optimized binaries. |
| SARD-TestVul-SDM | Os, Ofast | Assesses SDM in specific optimizations. | Validates versatility and accuracy. |
| SARD-CWE-HighOpt-SDM | O2, Ofast | Evaluates SDM's flexibility. | Shows adaptability and consistent detection. |
| SARD-CWE-LowOpt-SDM | O0, O1 | Challenges SDM with minimal optimization. | Ensures detection with minimal transformations. |
| CVE-Binaries-SDM | Various | Tests SDM on documented vulnerabilities. | Highlights precision in identifying vulnerabilities. |

- SARD-AdvOpt-SDM: This dataset focuses on binaries compiled with advanced optimization techniques, including O2 and O3 levels, from the SARD suite.
- SARD-TestVul-SDM: The success of the SDM to detect diverse vulnerabilities is the actual importance of this datasets. It assesses how well SDM detects vulnerabilities in code that has been optimized with GCC, mainly when using Os and Ofast optimizations.
- SARD-CWE-HighOpt-SDM: This dataset, which is similar to SARD-CVE-LowOpt, contains binaries optimized with O2 and Ofast. It plays a vital role in demonstrating how flexible the SDM is to code optimization of different levels and the effects of adherence to vulnerability detection.
- SARD-CWE-LowOpt-SDM: This dataset presents binaries constructed with O0 and O1 GCC optimization levels, presenting a challenge to the SDM in terms of normalizing differences brought about by variety of compiler optimizations and promising continuous detection of vulnerabilities.
- CVE-Binaries-SDM: Comprised of binaries with well-documented vulnerabilities, this dataset emphasizes the SDM's capacity to accurately capture the intricate patterns of known and unknown vulnerabilities, converting them into refined pseudo code for further analysis.

The Skip-Gram model [55], with a vector dimension of 100 and a window size of 5 constructs meaningful vector representations of these tokens. The SDM leverages this process to improve the semantic correlation between tokens, ensuring that the pseudo code generated provides a robust foundation for the subsequent vulnerability detection process. The selected datasets information is shown in *Table 3*.

*C. Neural Network Training*

We used TensorFlow and Keras to construct the model underpinned by SDM. SDM employs the combined strengths of IDA Pro 7.5 and the Hexray Decompiler [56], facilitated the extraction of high-quality ASTs from decompiled binaries. This robust feature extraction is critical for training our deep learning models.

Our neural network is initialized with zero vectors for the leaves of the AST, aligning with the focus on precision from the ground up. Training employs BCELoss for its suitability in binary classification, resonating with the separation of code features into "vulnerable" and "non-vulnerable."

Optimization is carried out using the AdaGrad algorithm, selected for its efficiency with sparse data— a common characteristic in the decompilation output. A carefully chosen training cycle of 30 epochs ensures a balance between underfitting and overfitting, a key consideration when using SDM-enhanced feature sets.

We experimented with various hidden embedding sizes and noted that while larger embeddings captured more complexity, a size of 256 hit an optimal balance for our SDM-augmented data, maximizing both efficiency and model performance.

In assessing the architecture, we discovered that a configuration of three BiLSTM layers processed the complexity of SDM-generated features effectively, with a token cutoff at 500 ensuring coverage of more extensive program structures without compromising model consistency. The parameters settings are shown in *Table 4*.

The neural network's parameters were accurately put to accommodate the rich feature set produced by SDM, with the computational resources detailed ensuring the model could process the augmented dataset without troubles.

The model operates within a Python 3.6 environment on a local server. The entire process, from dataset construction using buildroot-2020 [57], and GNU compiler collection v13.1 to model training set to capitalize on the benefits SDM introduces.

Table 4 Parameter of BiLSTM-Residual Attention

| Parameters | Values |
|---|---|
| Number of tokens in code gadget | 500 |
| Size of hidden embedding | 256 |
| Layers of BiLSTM | 3 |
| Batches sizes | 64 |
| Epochs | 30 |
| Optimizer | AdaMax |
| Learning rate | 0.001 |
| Dropout | 0.5 |

Table 5 Metrics formulas

| Metric | Formula |
|---|---|
| Accuracy | $A = \dfrac{TP + TN}{TP + TN + FP + FN}$ |
| Precision | $P = \dfrac{TP}{TP + FP}$ |
| F1-score | $\text{F1-score} = 2 \times \dfrac{P \times R}{P + R}$ |
| False-positive | $FPR = \dfrac{FP}{FP + TN}$ |
| False-negative | $FNR = \dfrac{FN}{TP + FN}$ |

*D. Measures of Evaluation*

In evaluating the performance of our vulnerability detection model, we utilize five standard metrics commonly applied in classification tasks. These are Accuracy, Precision, False Positive Rate (FPR), False Negative Rate (FNR), and the F1-score. Each metric is defined as follows: Accuracy (A) measures the proportion of true results (both true positives and true negatives) among the total number of cases examined. Precision (P) reflects the proportion of actual positive identifications which were correct, emphasizing the model's ability to minimize false positives. False Positive Rate (FPR) indicates the likelihood that non-vulnerable code gadgets are incorrectly classified as vulnerable. False Negative Rate (FNR) measures the proportion of vulnerabilities that the model fails to detect, with Recall being the complement of FNR, focusing on the model's sensitivity. F1-score (F1) is the harmonic mean of Precision and Recall, providing a balance between the model's precision and sensitivity. The mathematical representations for these metrics are standardized and crucial for the validation of our model's performance *Table 5* Evaluation metrics.

*E. Analysis and Results*

*Research experiments related to contribution 1:*

The experimental results on the **SARD-AdvOpt-SDM** dataset show that VulCatch outperforms other models, achieving an accuracy of 98.43%, precision of 97.92%, and an F1-score of 98.88%. Its slightly higher FPR of 1.56% compared to HAN-BSVD's 1.46% is offset by a lower FNR of 2.71%. VulCatch excels due to advanced synergy decompilation module and deep learning techniques, starting with CodeT5 to convert binary code into pseudocode, followed by detailed analysis using IDA and Ghidra. The Synergy Decompilation Module (SDM) and Kolmogorov-Arnold Networks (KAN) enhance feature extraction and dependency modeling. Inception Blocks, BiLSTM layers, and attention mechanisms with residuals further improved focus and learning efficiency. VulCatch outputs vulnerability scores with program slicing ensuring accuracy and effectiveness *Table 6* results of SARD-AdvOpt-SDM.

VulCatch achieves the highest accuracy at 98.43%, indicating its exceptional capability to correctly detect vulnerabilities, while HAN-BSVD with an accuracy of 97.66% also demonstrates strong performance. Other models like VulPin exhibits notable performance. VulCatch exhales likely due to its advanced feature extraction and deep learning techniques.

With a precision of 97.92%, VulCatch demonstrates its proficiency in minimizing false positives. This means that the vulnerabilities it detected are very likely to be actual vulnerabilities, reducing the time and resources wasted on false alarms. Although HAN-BSVD is nearly as precise with 97.90%, models like VulPin (93.48%) and BVDetector (89.23%) show a notable drop in precision, indicating a higher rate of false positives.

VulCatch's F1-score of 98.88% demonstrates its balanced recall and precision performance. This measure is essential since it guarantees that the model finds true positives accurately and precisely. HAN-BSVD, has a lower F1-score of 97.15%, reinforcing VulCatch's ability to handle both true positives and false positives effectively. Some model's performances like BVDetector (89.67%) and VulPin (94.25%) are less balanced, as seen by the considerable lags.

Although VulPin reaches a very low FPR of 0.15%, VulCatch is still very competitive with a low FPR of 1.56%. VulCatch hardly misclassifies non-vulnerabilities as vulnerabilities, as evidenced by this low false positive rate. The slightly higher FPR compared to VulPin suggests that while VulCatch might have a marginally higher rate of false positives, it compensates for this with overall better performance in other metrics. Models like Instruction2vec (7.43%) and Asteria-Pro (8.35%) exhibit much higher FPRs, indicating less reliability in avoiding false alarms.

VulPin again excels with the lowest FNR of 1.02%, followed closely by VulCatch at 2.71%. This low FNR for VulCatch means it effectively minimizes the number of actual vulnerabilities that are missed, which is critical for ensuring system security. Higher FNRs in models like Asteria-Pro

Table 6 Outcome of each method on SARD-AdvOpt-SDM datasets

| Model | A (%) | P (%) | F1 (%) | FPR (%) | FNR (%) |
|---|---|---|---|---|---|
| VulPin 2023, [58] | 95.98 | 93.48 | 94.25 | 0.15 | 1.02 |
| VDSimilar 2021, [59] | 89.14 | 87.99 | 87.22 | 2.26 | 6.94 |
| BVDetector 2020, [9] | 92.75 | 89.23 | 89.67 | 1.23 | 4.28 |
| Asteria-Pro 2023 [60] | 91.65 | 91.65 | 91.65 | 8.35 | 25.85 |
| VulDeePecker 2018, [27] | 87.66 | 84.49 | 85.13 | 11.69 | 13.17 |
| Instruction2vec 2019, [6] | 91.88 | 89.33 | 90.10 | 7.43 | 9.11 |
| HAN-BSVD 2021, [8] | 97.66 | 97.90 | 97.15 | 1.46 | 3.59 |
| **VulCatch** | **98.43** | **97.92** | **98.88** | **1.56** | **2.71** |

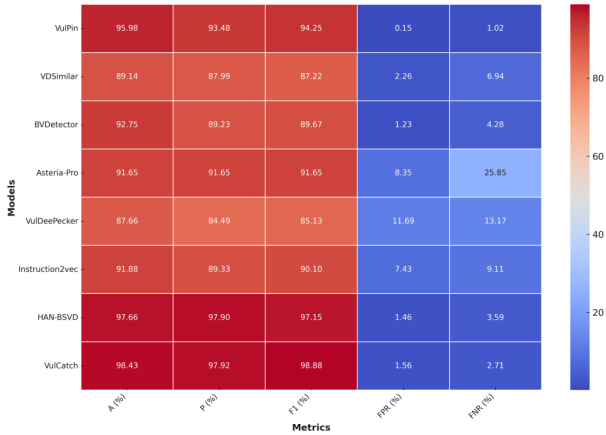

Fig. 4. heatmap outcome visualization on SARD-AdvOpt-SDM datasets

(25.85%) and VulDeePecker (13.17%) highlight their inactiveness in detecting true vulnerabilities.

The heatmap in *Fig. 4* provides a clearer visualization of the model's performance for quick comparisons. Each cell's color intensity indicates the magnitude of the metric, with darker shades representing higher values and lighter shades indicating lower values highlighting the relative strengths and weaknesses of each model. Models such as VulPin, HAN-BSVD, and VulCatch exhibit strong performance metrics overall, indicated by the darker shades in accuracy, precision, and F1 score columns and lighter shades in FPR and FNR columns. Meanwhile, models having higher FNR and FPR scores, e.g., VulDeePecker and Asteria-Pro, show possible areas for enhancement. After comparing the two models, it is possible to notice which is more reliable and efficient in terms of improving detection accuracy and reducing false positives and negatives. This heatmap is a helpful tool to measure and choose the models following their actual performance and strengths since it gives a thorough overview of the performance measures.

The lowest FPR and FNR is VulPin due to its simplified feature extraction and model architecture, it has a marginally lower accuracy (95.98%) and F1-score (94.25%). VDSimilar performance is worst according to all the metrics, most likely as a result of its outdated methodologies. BVDetector's performance is quite good but has lower accuracy (92.75%) and F1-score (89.67%) possibly due to less thorough feature extraction. High FPR (8.35%) and FNR (25.85%) are the problems with Asteria-Pro, which may be because of the less sophisticated deep learning layers and feature extraction techniques. VulDeePecker has lower accuracy (87.66%) and higher FPR (11.69%) and FNR (13.17%) due to its reliance on older methods. Instruction2vec achieves reasonable performance but does not incorporate the latest advancements, leading to slightly lower metrics. HAN-BSVD is a close competitor with high accuracy (97.66%) and precision (97.90%), though it has a slightly lower F1-score (97.15%) compared to VulCatch, likely due to less advanced methodologies. Compared to the models, VulCatch shows clear advantages.

**SARD-TestVul-SDM:** VulCatch achieves remarkable results with a high Accuracy 99.29, Precision 99.50, F1-score 99.05, and a low FPR 0.39%, FNR 1.37%. Compared to other baseline models like Bin2Vec, VulMatch, Gemini, i2v_attention, and SAFE, VulCatch consistently outperforms them in all measures *Table 7* outcome of SARD-TestVul-SDM. VulCatch demonstrates excellent detection results not only for single types of vulnerabilities but also for multiple types, showcasing its capability to simultaneously discover various kinds of vulnerabilities with just one trained model.

The comparison between systems like SAFE and i2v_attention, which extract semantic features, show better performance compared to systems extracting syntactic features like VYPER, Gemini, and VIVA. This suggests that semantic features contribute to better vulnerability detection.

VulCatch achieves the highest accuracy at 99.29%, demonstrating its ability to correctly detect vulnerabilities. This high accuracy underscores its robustness in handling diverse types of vulnerabilities within binary code. BinVulDet also performs well with an accuracy of 99.11%, closely trailing VulCatch. In comparison, systems like Gemini and VYPER have lower accuracies at 90.01%, and 78.4% respectively, showing their limit in efficiency.

Table 7 Outcome of each method on SARD-TestVul-SDM datasets

| Model | A (%) | P (%) | F1 (%) | FPR (%) | FNR (%) |
|---|---|---|---|---|---|
| VIVA 2021, [10] | 92.4 | 87.6 | 88.4 | 14.6 | 9.4 |
| BinVulDet 2022, [23] | 99.11 | 99.25 | 98.83 | 0.46 | 1.59 |
| VYPER 2020, [1] | 78.4 | 57.4 | 57.42 | 0.6 | 42.6 |
| PDiff 2020, [39] | 95.87 | 94.20 | 95.04 | 2.00 | 3.96 |
| Gemini 2017, [34] | 90.01 | 68.96 | 58.80 | 3.73 | 48.76 |
| i2v_attention 2019, [61] | 95.6 | 96.1 | 95.9 | 4.4 | 4.4 |
| SAFE 2021, [4] | 91.87 | 79.60 | 62.75 | 2.02 | 48.22 |
| **VulCatch** | **99.29** | **99.50** | **99.05** | **0.39** | **1.37** |

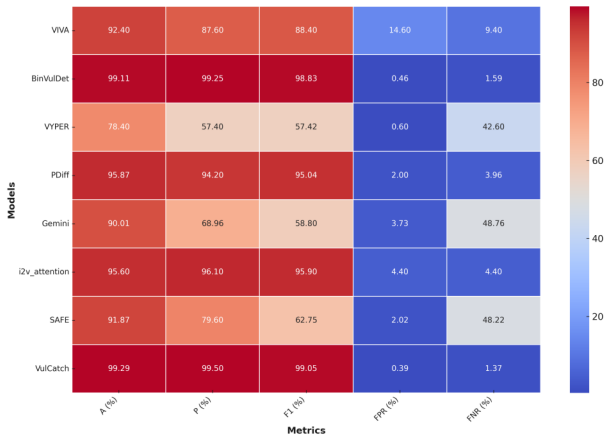

Fig 5. heatmap outcome visualization on SARD-TestVul-SDM datasets

VulCatch's precision of 99.50% shows its high competence in minimizing false positives. It means the vulnerabilities detected are highly likely to be real vulnerabilities, which is important in false alarms reduction. BinVulDet is the second, having a precision of 99.25%, but Gemini and VYPER have much lower precision of 68.96%, and 57.4% respectively, showing their higher false positives rates.

The F1-score of 99.05% for VulCatch indicates its balanced performance in recall and precision. This is a crucial metric that ensures VulCatch not only is precise but also effective in detecting true positives. BinVulDet slightly has lower F1-score of 98.83%, which indicates solid but marginally less balanced performance. Models such as Gemini and SAFE have much lower F1-scores of 58.80%, and 62.75% respectively, indicating a less balance between recall and precision.

VulCatch has maintained a very low FPR of 0.39%, signifying that it hardly misclassifies non-vulnerabilities as vulnerabilities. This low FPR is important for decreasing false alarms in real-world applications. BinVulDet has a marginally higher FPR of 0.46%, while models such as i2v_attention and VIVA have much higher FPRs of 4.4%, and 14.6% respectively, signifying a higher false positives rate.

VulCatch attains a low FNR of 1.37%, indicating its efficiency in decreasing missed vulnerabilities. The low FNR is important to make sure actual vulnerabilities are not overlooked, improving the general security posture. BinVulDet has a slightly higher FNR of 1.59%. On the other hand, VYPER and SAFE struggle significantly with FNRs of 42.6% and 48.22%, respectively, indicating they miss a large number of actual vulnerabilities.

The heatmap in *Fig 5* provides a visual comparison of each metrics. Each cell's color intensity represents the value of the metric, with darker colors indicating higher values and lighter colors indicating lower values for quick identification of patterns and outliers among the models. It also highlights the relative strengths and weaknesses of each model. Models such as BinVulDet and VulCatch demonstrate strong overall performance, as evidenced by the dark shades in accuracy, precision, and F1 score columns, and lighter shades in the FPR and FNR columns. On the other hand, models like SAFE and VYPER exhibit higher false negative rates, impacting their overall performance negatively. This comparative analysis helps in understanding which models are more reliable and efficient.

The comparisons between models like i2v_attention and SAFE which focuses on semantic features extraction, exhibits better performance than models extracting syntactic features like VIVA, VYPER, and Gemini. This indicates that, semantic features usage to capture deeper and more meaningful patterns in the data, contributes significantly to improved vulnerability detection. Integrating deep learning techniques and advanced decompilation in VulCatch, alongside its capability to extract and utilize semantic features effectively result in better outcomes.

The Test Vulnerability (TestVul) and Advanced Optimization (AdvOpt) datasets were used because of their ability to deliver realistic and comprehensive evaluations. The complexity of AdvOpt's dataset and practicality mirrors real-world scenario with obfuscated and optimized code, allows rigorous testing of models' abilities across different vulnerabilities. Its benchmarking status allows direct comparisons with current models, it highlights advanced methods like VulCatch's Synergy Decompilation Module and Kolmogorov-Arnold Networks. The TestVul dataset focuses on both known and unknown vulnerabilities, presenting a controlled, challenging settings which includes a diverse vulnerability type for comprehensive evaluation. This consistent benchmarking confirms the reliability and comparability for evaluation metrics related to real-world applications. Together, these datasets validate VulCatch's greater performance in both optimized, advanced environments, and real-world situations.

Table 8 Outcome of each method on SARD-CWE-HighOpt-SDM datasets

| Models | BiLSTM-Attention | BiLSTM | GRU | LSTM | BiGRU | BiGRU-Attention | Transformer | Our Model |
|---|---|---|---|---|---|---|---|---|
| FPR (%) | 0.46 | 0.61 | 0.71 | 0.76 | 0.49 | 0.92 | 0.95 | **0.46** |
| FNR (%) | 1.59 | 1.73 | 1.49 | 1.68 | 1.72 | 1.41 | 10.12 | **1.40** |
| A (%) | 99.12 | 98.96 | 98.99 | 98.89 | 99.01 | 98.89 | 95.55 | **99.19** |
| P (%) | 99.26 | 99.00 | 98.85 | 98.76 | 99.21 | 98.51 | 98.31 | **99.45** |
| F1 (%) | 98.84 | 98.63 | 98.68 | 98.54 | 98.74 | 98.55 | 93.91 | **98.90** |

Table 9 Outcome of each method on SARD-CWE-LowOpt-SDM datasets

| Models | BiLSTM-attention | BiLSTM | GRU | LSTM | BiGRU | BiGRU-attention | Transformer | Our Model |
|---|---|---|---|---|---|---|---|---|
| FPR (%) | 1.78 | 1.09 | 0.99 | 0.97 | 0.49 | 2.07 | 3.69 | **0.39** |
| FNR (%) | 1.19 | 2.06 | 1.93 | 3.60 | 3.28 | 1.10 | 8.74 | **1.37** |
| A (%) | 98.41 | 98.60 | 98.71 | 98.20 | 98.63 | 98.24 | 94.72 | **98.79** |
| P (%) | 96.22 | 97.62 | 97.83 | 97.86 | 98.90 | 95.64 | 91.90 | **99.00** |
| F1 (%) | 97.50 | 97.78 | 97.95 | 97.12 | 97.80 | 97.24 | 91.58 | **98.88** |

*Research experiments related to contribution 2:*

SARD-CWE-HighOpt-SDM and SARD-CWE-LowOpt-SDM datasets extensive experiments shown important insights into the performance of different neural network models for vulnerability detection. We used a collection of neural network models, involving LSTM, GRU, BiLSTM, BiGRU, their attention-augmented versions, and the Transformer model. Moreover, we presented an innovative BiLSTM-R-A model, which showed outstanding capability in these tasks.

Our proposed model achieved outstanding results on the SARD-CWE-HighOpt-SDM dataset, overtaking all other models except for same FPR of BiLSTM Attention model. It recorded a remarkable accuracy rate of 99.19%, the highest precision of 99.45%, and an impressive F1 score of 98.90%, while maintaining an FPR of 0.46% and a competitive FNR of 1.40%. The comparative analysis indicates that bidirectional networks like BiGRU and BiLSTM generally decreased the FPR by an average of 0.24% relative to unidirectional networks. The integration of attention mechanisms with bidirectional networks proved powerful, as demonstrated by the BiLSTM-attention and BiGRU-attention models in Table *8* results of SARD-CWE-HighOpt-SDM.

On the SARD-CWE-LowOpt-SDM dataset, despite traditional Transformer models being less effective, bi-directional neural networks enhanced with attention mechanisms, such as BiLSTM and BiGRU, demonstrated reasonable competence. Our model excelled in this dataset as well, achieving lower FPR and FNR, and higher accuracy, precision, and F1 scores compared to other models. Specifically, our model achieved the lowest FPR of 0.39%, a competitive FNR of 1.37%, the highest accuracy of 98.79%, precision of 99.00%, and an F1 score of 98.88% in *Table 9* results of SARD-CWE-LowOpt-SDM. These results underscore the significance of incorporating advanced bi-directional and attention-based architectures.

The differing performances of the same models on the SARD-CWE-HighOpt-SDM and SARD-CWE-LowOpt-SDM datasets are due to the variations in the quality of pseudo-code obtained from decompilation. The HighOpt dataset contains higher-quality, better-structured code, making it easier for models to learn patterns and dependencies, resulting in higher accuracy and lower false positive and false negative rates. In comparison, the LowOpt dataset has lower-quality code with more errors and ambiguities, posing challenges for the models. This leads to higher false positive and false negative rates and generally lower performance metrics. The impact is mainly obvious in models such as Transformer, which relies on global attention mechanisms that are more effective with high-quality data. Models improved with bidirectional and attention mechanisms, such as BiGRU-Attention and BiLSTM-Attention, perform quite better on the LowOpt dataset by efficiently catching the context and concentrating on important portions of the data. Our proposed model's consistent performance across both datasets shows its adaptability and robustness.

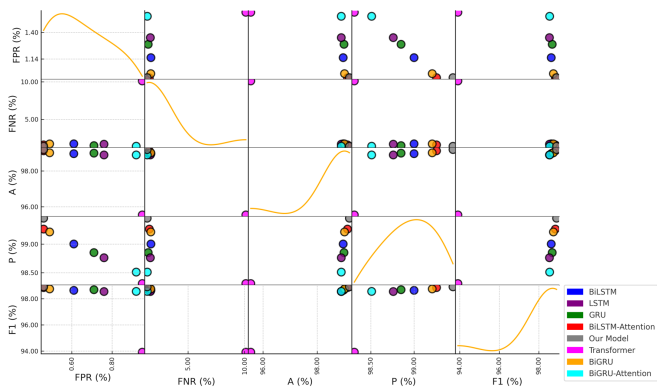

Fig. 6. Outcome of each method on SARD-CWE-HighOpt-SDM datasets

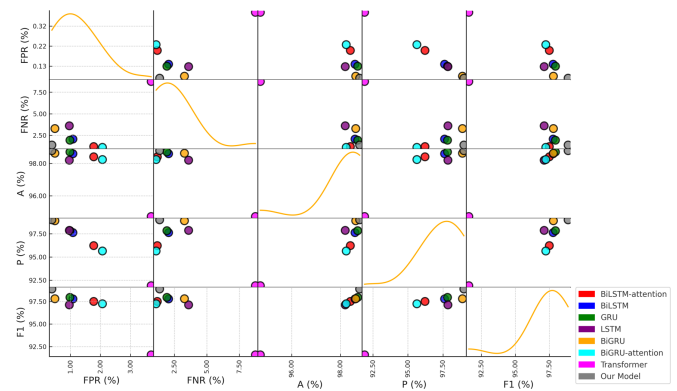

Fig. 7. Results of SARD-CWE-LowOpt-SDM neural networks

Comparing the HighOpt and LowOpt datasets is crucial for evaluating the robustness and adaptability of neural network models in real-world security applications. This comparison allows us to assess how well models can maintain their performance across varying data quality levels, ensuring they are not only effective in optimal conditions but also resilient and reliable when faced with the imperfections typical of practical decompilation processes. By demonstrating consistent high performance across both datasets, a model proves its robustness and practical applicability in diverse and unpredictable security contexts.

The scatter matrices visually represent the relationships between all performance metrics for all the models. Diagonal elements are histograms showing the distribution of each metric, indicating how frequently different values occur. For instance, most models have a low FPR, with "Our Model" having the lowest at 0.39%. Similarly, high accuracy values are clustered between 94% and 99%, with "Our Model" achieving the highest accuracy at 98.79%.

Off-diagonal elements are scatter plots that show pairwise relationships between metrics, with annotated model names. These plots help identify correlations or trade-offs between metrics. A plot shows that models with lower false negative rates tend to have higher accuracy. Patterns in these scatter plots provide insights into model performance. These scatter matrices help identify key performance trends and outliers. "Our Model" stands out with low FPR and FNR, and high accuracy, precision, and F1 scores, indicating superior performance. In contrast, "Transformer" shows a high FNR, affecting its overall accuracy. These visualizations allow for informed comparisons and evaluations of each model's strengths and weaknesses in *Fig. 6* and *Fig. 7* result of SARD-CWE-HighOpt-SDM & SARD-CWE-LowOpt-SDM.

Overall, the synergy between bidirectional neural networks and attention mechanisms, especially embodied by the BiLSTM-attention model, significantly enhances performance in vulnerability detection tasks. Our proposed model emerges as a classic standard, seamlessly integrating these elements to achieve superior detection rates. Its exceptional performance across the SARD-CWE-HighOpt-SDM and SARD-CWE-LowOpt-SDM datasets highlights its adaptability and robustness, setting a new benchmark for automated vulnerability detection systems.

This model not only excels in detecting vulnerabilities with greater accuracy but also demonstrates adaptability for SDM, positioning it as a remarkable tool for vulnerability detection.

*Research experiments related to contribution **3**:*

In the evaluation of VulCatch, we observed its performance across seven distinct CVEs representing a wide range of vulnerabilities from different years and domains in *Table 10* CVE data outcomes. The SDM has proven effective, with the deep learning model successfully capturing hierarchical and contextual information through multi-level feature extraction.

**CVE-2018-1002105** is related to Kubernetes API, VulCatch correctly detected 15 out of 16 code gadgets, with just one inaccurate detection. This outcome shows the robustness of the model in handling complex server-side vulnerabilities comprising of privilege escalation.

With **CVE-2019-0708** (BlueKeep), a notorious remote code execution vulnerability, VulCatch attained accuracy even higher, with a correct detection of 20 out of 22 code gadgets. The ability of the model to understand remote desktop protocol details is clear in these results.

The model was tested with **CVE-2017-5715** (Spectre) and **CVE-2017-5754** (Meltdown) for its aptitude of speculative execution vulnerabilities. Handling Spectre with an accuracy of 12 accurate detections and mitigated Meltdown with 100% accuracy, representing a clear understanding of processor-level operations.

For Windows VBScript engine vulnerability **CVE-2018-8174**, VulCatch's had a robust performance, accurately detecting 17 code gadgets, with only three inaccuracies, showing efficiency in script-based vulnerability detection.

Table 10 Results of CVE datasets

| CVE Identifier | Description | Accurate | Inaccurate | Code Gadgets |
|---|---|---|---|---|
| CVE-2018-1002105 | Kubernetes API server vulnerability allowing privilege escalation | 15 | 1 | 10 |
| CVE-2019-0708 | Remote Desktop Services Remote Code Execution Vulnerability, known as BlueKeep | 20 | 2 | 12 |
| CVE-2017-5715 | Systems with microprocessors utilizing speculative execution and indirect branch prediction may allow unauthorized disclosure of information to an attacker | 12 | 2 | 8 |
| CVE-2017-5754 | Systems with microprocessors utilizing speculative execution and indirect branch prediction may allow unauthorized disclosure of information to an attacker | 18 | 0 | 9 |
| CVE-2018-8174 | Remote code execution vulnerability in Windows VBScript engine | 17 | 3 | 11 |
| CVE-2020-0601 | A spoofing vulnerability in Windows CryptoAPI reported by the NSA, known as CurveBall or ChainOfFools | 19 | 1 | 10 |
| CVE-2020-1472 | An elevation of privilege vulnerability in Microsoft Netlogon, known as Zerologon | 16 | 2 | 7 |

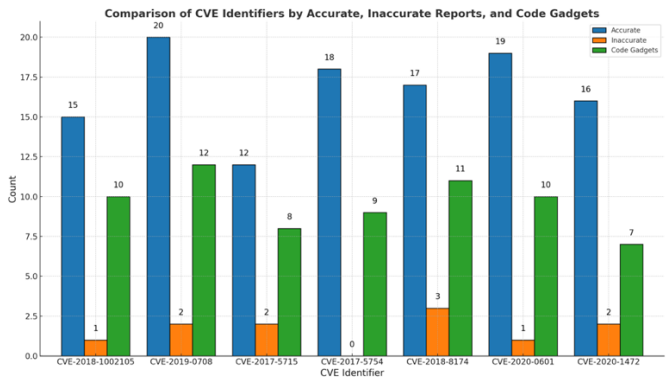

Fig. 8. Results of CVE datasets

**CVE-2020-0601** (CurveBall/ChainOfFools) usage involved cryptographic complications, but VulCatch kept high precision of 19 accurate detections, demonstrating its competence in the cryptographic area.

Finally, the model was evaluated against **CVE-2020-1472** (Zerologon), a complex elevation of privilege vulnerability, it achieves a success rate of detecting 16 out of 18 code gadgets accurately, all shown in CVE data outcomes *Fig. 8*.

We noticed cases where reported accuracy of vulnerability reports exceeded the amount of detected code gadgets. This difference stems from the multifaceted nature of vulnerability reporting; multiple accurate reports accompanying a single CVE, but code gadgets signify tangible vulnerabilities in the software.

Performance and comparison analysis and of VulCatch across different parts:

- Accuracy in Different Domains: VulCatch displayed high accuracy across vulnerabilities stemming from network operations, speculative execution, and cryptographic vulnerabilities. This suggests the tool's wide applicability.
- False Positive and False Negative Rates: VulCatch showed a low rate of false positive, as witnessed by the smooth detection in certain cases. However, some minor discrepancies in complex vulnerability detections signify areas for enhancement.
- Complex Vulnerabilities Handling Capabilities: VulCatch was able to keep high rates of detection for sophisticated security problems like the Zerologon, confirming its potential.
- Extraction Hierarchical and Sequential Feature: BiLSTM-R-A was key in forcing the model to train all the data including the leftovers, this results in high accurate detection rates.
- Vulnerability Generalization across Software and Hardware: VulCatch has confirmed its flexibility by generalizing across both software and hardware vulnerabilities, proving the system's flexibility.
- Performance and Real-world Applicability: With fine-tuning for generalization and accuracy, VulCatch has showed potential in real-world settings, proving its usefulness beyond laboratory settings.
- Security Practices Implications: The performance of VulCatch indicates significant implications for proactive security practices, stressing the importance of reliable detection system in managing vulnerabilities.

Demonstrates impressive performance in VulCatch with continuing improvement, mainly in script-based vulnerability detection, is important for the growing area of cybersecurity. The low false positive and false negative rates are sign of a mature system, however, there is always room for improvement, especially with the increase of new vulnerabilities and exploitation methods. This analysis underscores the need for continuous learning and adaptation in vulnerability detection systems.

*F. Overview*

The VulCatch program is evaluated based on the time it takes for three main tasks: preprocessing, training, and detection. **Preprocessing**: VulCatch took about 380 seconds to construct and encode 843 gadgets from 564 functions, averaging around 581 milliseconds per gadget. **Training**: For training, VulCatch utilized a dataset comprising 80% of the SDM data, with the neural network training taking nearly 309,000 seconds across 50 epochs (around 6200 seconds per epoch) and requiring over 3 million parameters. **Detection**: In the detection phase, VulCatch took 245 seconds to process 980 gadgets, which are about 483 milliseconds per gadget.

Feature Enhancement: By transforming the aggregated data into a more robust feature set, KAN enhances the subsequent analytical processes within VulCatch. The enhanced features allow for more accurate generation of ASTs and PDGs, which are critical for the precise classification of vulnerabilities. Generalization and Over fitting: KAN's ability to generalize from training data to unseen data is a key advantage, especially in the dynamically changing field of cybersecurity. Furthermore, the structure of KAN inherently mitigates the risk of overfitting, a common challenge in deep learning models dealing with complex datasets. In a comparative analysis with other models using the SARD dataset, it was noted that larger sample sizes led to longer preprocessing and detection times. VulCatch generally required more time during these phases due to its thorough analysis, which includes program slicing, advanced decompilation, and embedding structural and sequential information. It detects code gadgets instead of functions, necessitating more iterations and time. Despite this, the improved detection efficiency makes the time investment for VulCatch worthwhile.

V. CONCLUSION

In this paper, we introduced VulCatch, an innovative system designed to enhance the detection of vulnerabilities in binary code through a novel Synergy Decompilation Module (SDM) and Kolmogorov-Arnold Networks (KAN), combining advanced decompilation techniques and deep learning models. The system uses CodeT5 for initial decompilation, transforming binary code into pseudocode to retain high-level semantics information for comprehensive analysis. The pseudocode is analyzed further with traditional decompilers IDA and Ghidra, ensuring a comprehensive examination of possible vulnerabilities. Kolmogorov-Arnold Networks (KAN) is integrated for feature sets enhancement gotten from traditional decompilers and CodeT5 combined data, resulting in the generation of Abstract Syntax Trees (ASTs) and Program Dependency Graphs (PDGs), allowing Inception Blocks for multi-level feature extraction from these vectors. These components are important in understanding the hierarchical and dependency parts of the codes, helping pinpoint the critical parts more likely to have vulnerabilities. The data analysis from ASTs and PDGs employs a BiLSTM neural network comprising of attention and residual mechanisms, enhancing the detection efficiency and accuracy. Experimental outcome confirms the robustness of the model across seven CVE datasets, significantly increases accuracy and precision, decreasing false positive and negative rates. This makes VulCatch a trustworthy system for detecting both known and unknown vulnerabilities in real time.

VulCatch model represents a major improvement in the field of binary vulnerability detection. By the combination of traditional decompilers with modern deep learning methods, VulCatch addresses the limitations of current methods, offering a more comprehensive and accurate detection system. The use of CodeT5 for decompilation confirms the semantic integrity of the code is maintained, which is important for accurate detection of vulnerabilies. The experimental outcome validates the efficiency of VulCatch through different types and optimization levels of vulnerabilities. The ability of the system in handling binaries with different optimization levels proves its adaptability and robustness, making it a valuable system for real-world settings where binary code can vary significantly. However, the system's complication and the computational resources needed for training the deep learning models poses difficulties. The need for large datasets to train the system also underlines the importance of having access to diverse and comprehensive binary datasets.

VulCatch offers a major advancement in binary vulnerability detection but has some limitations. The deep learning system need extensive computational resources, which can be an obstacle for some organizations. The system's efficiency depends on diverse and extensive datasets that are difficult to obtain. The complication of incorporating multiple components, such as CodeT5 and traditional decompilers, makes maintenance hard. Furthermore, while the model performs good on tested datasets, its generalizability to less common or highly specialized binary code remains uncertain. To tackle these limitations, works in the future will focus on decreasing computational requirements through more hardware optimizations and efficient algorithms. Efforts will be made to diversify and enlarge datasets to increase generalizability and robustness. visualization tools, enhancing feature engineering methods and developing user-friendly interfaces will help in remediation and vulnerability analysis. Integrating VulCatch with current security system will enable exploring advanced models such as transformer-based architectures and wider adoption, could improve detection efficiency and accuracy further. These stages aim to improve VulCatch's capability to detect both known and unknown vulnerabilities in a wider range of binary code scenarios.

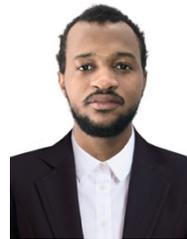

**Abdulrahman Chukkol** received the BSc Information Technology from NIMS University Jaipur, India in 2015, received MSc Information and Communication Engineering from Beijing Institute of Technology in 2021, and is currently pursuing PhD at Beijing Institute of Technology, his current research interests include Software security, Vulnerability Detection, Deep learning and neural networks.

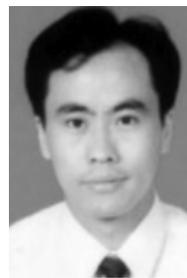

**Senlin Luo** received a Ph.D. in computer science from the Beijing Institute of Technology. He is currently the deputy director, laboratory director, and a professor with the Information System & Security and Countermeasures Experimental Center, Beijing Institute of Technology, Beijing 100081, China. His current research interests include machine learning, medical data mining, and information security.

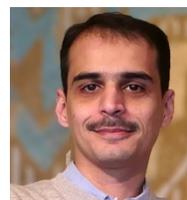

**Kashif Sharif** received his M.S. degree in information technology and Ph.D. degree in computing and informatics from the University of North Carolina at Charlotte in 2004 and 2012, respectively. He is currently an associate professor at Beijing Institute of Technology. His current research interests include wireless and sensor networks, network simulation systems, software defined and data center networking, ICN, and the Internet of Things. He is a member of ACM.

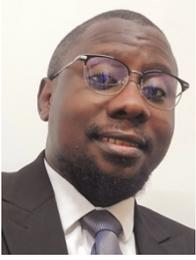
**Yunusa Haruna** receives his PhD from Beihang University in China, his research interests include Pattern recognition and intelligent system, object detection, convolutional neural network, and robust computer vision algorithms.

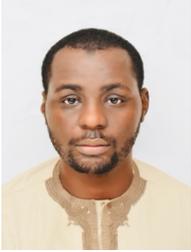
**Muhammad Abdullahi** received the BSc Computer schience management form Ism Adonia University in Benin, he is currently a Masters student at American University of Nigeria, his research interests include, Software engineering, Computer security and countermeasures.